\documentclass[preprint]{aastex}

\begin{document}

\title{The BeppoSAX 0.1 - 100 keV Spectrum of the X-ray Pulsar 
4U~1538--52}

\author{N.R. Robba\altaffilmark{1}, 
L. Burderi\altaffilmark{2},
T. Di Salvo\altaffilmark{3,1},
R. Iaria\altaffilmark{1},
G. Cusumano\altaffilmark{4}
}
\altaffiltext{1}
{Dipartimento di Scienze Fisiche ed Astronomiche, Universit\`a di Palermo, via 
Archirafi 36 - 90123 Palermo, Italy}
\authoremail{robba@gifco.fisica.unipa.it}
\altaffiltext{2}{Osservatorio Astronomico di Roma, Via Frascati 33, 
00040 Monteporzio Catone (Roma), Italy}
\altaffiltext{3}{Astronomical Institute "Anton Pannekoek," University of 
Amsterdam and Center for High-Energy Astrophysics,
Kruislaan 403, NL 1098 SJ Amsterdam, the Netherlands}
\altaffiltext{4}{Istituto di Fisica Cosmica con Applicazioni 
all'Informatica CNR, Via U. La Malfa 153, Palermo, I-90146, Italy}

\begin{abstract}
We report  the results of temporal  and spectral analysis performed on
the X-ray pulsar 4U~1538--52 observed by BeppoSAX.   We obtained a new
estimate of the spin period of the neutron  star $P=528.24 \pm 0.01$~s
(corrected for the orbital motion of  the X-ray source): the source is
still  in  the spin-up state,  as since  1988.  The  pulse  profile is
double peaked,   although   significant  variations of   the  relative
intensity of  the peaks   with energy  are  present.  The  broad  band
(0.12--100~keV)  out-of-eclipse  spectrum   is well  described   by an
absorbed  power law modified by a  high energy cutoff at $\sim 16$~keV
({\it e}-folding energy $\sim  10$~keV) plus an  iron emission line at
$\sim 6.4$~keV.  A  cyclotron line at  $\sim 21$~keV is present.   The
width of the line is consistent with thermal Doppler broadening at the
temperature  of the exponential cutoff.   We searched for the presence
of the second harmonic, previously reported  for this source. We found
no evidence of lines at $\sim  42$~keV, although an absorption feature
at 51~keV seems  to  be present (at   99\% confidence level).  A  soft
excess, modelled by a blackbody with a  temperature of $\sim 0.08$~keV
could be present, probably emitted by the matter at the magnetosphere.
We also performed  a spectral analysis  during the X-ray eclipse.  The
spectral evolution   during  the eclipse  can be  well  described by a
progressive covering  of  the primary Comptonization  spectrum that is
scattered   into the line  of sight.    During the  deep  eclipse this
spectrum  also softens,  suggesting that the  dust-scattered component
becomes important.  An alternative, more complex model,  with an emission  
iron line and scattered components  (as the one that has been used  to fit 
the eclipse  of Centaurus X-3), also gives a good  fit   of  the 
deep-eclipse data.
\end{abstract}


\section{Introduction}

4U~1538--52 is a wind-fed X-ray binary system formed by a massive 
($17\; M_{\odot}$) B0 star and a neutron star spinning with a period 
of about 529~s  (Davison 1977; Becker {\it et al.} 1977). 
The X-ray luminosity has been estimated  $\sim 2 \times 10^{36}$~ergs/s
for a distance of $\sim 6.4$~kpc (Becker {\it et al.} 1977). 
The orbit, in an almost edge-on plane, is characterized by a period of 
3.75~days, a low eccentricity of $\sim 0.08$, and
a well defined X-ray eclipse lasting $\sim 0.6$~days
(Davison 1977; Becker {\it et al.} 1977).
Pulse period measurements show that the neutron star was in a spin-down
state before 1988, with a $|\dot{P}/P| \sim 10^{-11}$~s$^{-1}$, 
and in a spin-up state with the same $|\dot{P}/P|$ after 1988 
(Rubin  {\it et al.} 1997).

Before the first Ginga observation (Clark {\it et al.} 1990) the X-ray 
spectrum of 4U~1538--52 has been well modeled by a power-law modified by a
high-energy exponential cutoff and an iron emission line at 6.7~keV
(equivalent width $EW \sim 100$~eV).  A phase-dependent absorption
feature around 20~keV was observed in the X-ray spectrum by {\it
Ginga} (Clark {\it et al.} 1990), and was explained as cyclotron
resonance absorption.  A cutoff in the spectrum starting at $\sim
30$~keV was interpreted as the second harmonic but the {\it Ginga}
energy range precluded any definitive conclusion.  A fit of the 
energy-dependent pulse profiles in the range 10--38~keV to theoretical models
of the pulsar emission has been performed by Bulik {\it et al.} (1992,
1995).  The results are in agreement with pencil emission from a
strongly magnetized optically thick ($\tau\sim 20$) short ($h\ll
R_{NS}$) slab of plasma, heated by Coulomb collision stopping
mechanism in the accreting matter, in line with theoretical
calculations of Meszaros \& Nagel (1985a,b) and Brainerd \& Meszaros
(1991) for the non-relativistic and the relativistic case,
respectively.  More recently Makishima {\it et al.} (1999), in a paper
discussing {\it Ginga} observations of a number of X-ray pulsators,
briefly reported the results of an analysis of a 3 -- 60 keV {\it
Ginga} spectrum acquired during an observation on 1991 July 27. 
They fitted
the continuum with the so-called NPEX model (Mihara 1995) that is the
sum of two power--laws, of negative and positive index respectively,
both multiplied by an exponential cutoff of thermal origin. A
cyclotron feature at $\sim 20$ keV was fitted by a pseudo--Lorentzian
shape (i.e.\ a Lorentzian multipled by the energy squared, Mihara 1995). 
With this model they obtained a value of the $\chi^2/$d.o.f. of $46/29$.

In this paper we present for the first time a broad band (0.12--100 keV)
spectrum of 4U~1538--52. We clearly show that this spectrum can be well
fitted only if a thermal cutoff and a cyclotron absorption feature at
$\sim 21$ keV are included in the model. 
We also do not find strong evidence of a second harmonic at $\sim 42$ 
keV, although a weak feature could be present at a slightly higher energy.

\section{Observations and Temporal Analysis}

BeppoSAX observed 4U~1538--52 with its Narrow Field Instruments 
(NFI, Boella {\it et al.} 1997) in 1998 from July 29 
to August 1.
The NFIs are four co-aligned instruments with a broad band coverage 
from 0.1~keV up to 200~keV, with good
spectral resolution in the whole range. The Low Energy Concentrator 
Spectrometer (LECS, energy range 0.1--10~keV) and the Medium Energy
Concentrator Spectrometer (MECS, 1--11~keV) have
imaging capabilities with a field of view (FOV) of $20'$ and $30'$ radii, 
respectively. We selected the data for scientific analysis 
in circular regions of the FOV, centered on the source, 
of $8'$ and $4'$ radii for 
LECS and MECS, respectively. The background subtraction is usually
obtained using blank-sky observations, in which the background spectra 
are extracted from regions of the FOV similar to those used for 
the source. Because of the possibility of a contribution from the Galactic 
ridge at the coordinates of 4U~1538--52, we used
a particular attention to the background estimation. This is particularly
important for the spectral analysis during the eclipse, when the source
flux is at a very low level.
Local background has been therefore measured in a region of the image 
$16'$ and $20'$ far from the source for LECS and MECS, respectively, and 
compared with the background of the archive blank fields. LECS and MECS
local background spectrum is compatible with the blank-field spectrum, and 
therefore blank-sky background, which has a better statistics, has been used.
The High Pressure Gas Scintillator Proportional Counter (HPGSPC, 7--60~keV) 
and the Phoswich Detector System (PDS, 13--300~keV) have not imaging 
capabilities, and their FOVs, of $\sim 1^\circ$ FWHM, are 
delimited by collimators. The background subtraction for these instruments is 
obtained using off-source data accumulated during the rocking 
of the collimators.   
The energy ranges used in the spectral analysis for the NFIs 
are: 0.12--4~keV for the LECS, 1.8--10.5~keV
for the MECS, 8--30~keV for the HPGSPC, and 15--100~keV for the PDS.
To take into account intercalibration systematics, 
different normalizations of the NFIs are considered in the spectral 
analysis by including normalizing factors, fixed to 1 for the 
MECS, and kept free for the other instruments. 
A systematic error of 1\% was added to all the spectra, although, for this
relatively weak flux, it is negligible when compared to the statistical
error.

Figure 1 shows the 4U~1538--52 light curve, binned at 5700 s, in four 
different energy bands: 
0.1--2.5~keV (LECS data), 3--5~keV (MECS data), 5--10~keV (MECS data),
15--200~keV (PDS data).
This light curve shows the source in its eclipse state during the first 
$\sim 52$ ks and out of eclipse in the last $\sim 177$ ks.
A dip, probably due to low energy absorption, is also present in the 
0.1--2.5 keV light curve.
Adopting a distance of 6.4 kpc (Becker {\it et al.}, 1977) and
the spectral model described in the next section (see Tab.~1), we derived
the source luminosity out of eclipse, that is 
$\sim 1.8 \times 10^{36}$ ergs/s in the energy range 2--10~keV and 
$\sim 4.7 \times 10^{36}$ ergs/s in the whole range 0.1--100~keV. 

We performed a temporal analysis on the out-of-eclipse data
to search for the pulse period of the source using MECS data, 
which have the best statistics. 
The arrival times of all the events were reported to the solar system 
barycentre.  We corrected the arrival times for the orbital
motion of the source adopting the orbital parameters 
reported by Rubin {\it et al.} (1997), $P_{orb} = 3.72839 \pm 0.00002$ days
and $a \sin i = 53.5 \pm 1.4$ lt-s, and neglecting the eccentricity 
because of its low value. From the MECS data we estimated 
that the center-of-eclipse time is $T_{ecl} = 51024.460 \pm 0.032$ MJD. 
With these values we corrected the event times according to the formula:
$$
\Delta t = (a \sin i) \cos[2 \pi (t-T_{ecl})/P_{orb}].
$$
Then we performed a folding search for the best pulse period on these 
corrected arrival times. The best period obtained was $528.24 \pm 0.01$ s, 
demonstrating that the source is still in a spin-up state.
We checked the orbital correction {\it a posteriori} dividing 
the whole data set into consecutive intervals, which we folded using
the spin period derived above. No phase shifts were observed between
these intervals.

Figure 2 shows the pulse profiles during out-of-eclipse intervals in different 
energy bands, namely 0.1-1.8~keV (LECS, upper panel), 1.8-10.5~keV (MECS, 
middle panel), 15-100~keV (PDS, lower panel). 
A double peaked pulse profile is present in the 1.8-10~keV band.
The secondary peak is asymmetric in this energy range, and the
primary peak is broad and shows some structures at its maximum.
The secondary peak is absent in the soft range 0.1-1.8~keV,
while the primary peak still shows notch-like structures
at its maximum.
In the PDS range (15-100~keV) the emission is still pulsed, 
with the secondary peak less pronounced and more symmetric
and the primary peak narrower and smooth.
We have also produced PDS folded light curves in the following energy ranges:
15--20 keV (below the cyclotron line energy, see Clark et al. 1990 and
below in this paper), 20--30 keV (around the cyclotron line energy), and
30--100 keV (above the cyclotron line energy). In agreement with Ginga
results (Clark et al. 1990) we find that the secondary peak  
disappears at the cyclotron energy. Because of the low statistics,
there is no evidence that the secondary peak re-appears at energies 
higher than 30 keV.
In Figure 3 we plotted the MECS folded light curves in the energy ranges
1.8-5~keV and 5-10.5~keV and their hardness ratio. Several structures are
visible in this ratio, demonstrating the high spectral variability of 
the emission with the pulse phase.

\section{Spectral analysis}

\subsection{Pulse Phase Averaged Spectrum}

We performed spectral analysis on the post-egress energy spectrum 
of 4U~1538--52 in the energy range 0.12-100~keV. 
The high energy part of the spectrum (above $\sim 10$~keV)
is not well fitted by a power law modified by two absorption features 
as proposed by Clark {\it et al.} (1990) in the analysis of {\it Ginga} data. 
The broader band of the BeppoSAX instruments unambiguously demonstrates that
an absorption feature and an exponential cutoff are needed 
to adequately model the spectrum.

We fit the spectrum with an absorbed power-law continuum modified  
by a high energy cutoff and by an absorption cyclotron line 
of Gaussian shape plus a Gaussian emission line at $\sim 6.4$~keV
due to fluorescence of iron in low ionization stages.
The model used to fit the cyclotron line is a multiplicative Gaussian
of the form: $1 - {\rm Depth} \exp[-(E - E_{cyc})^2/(2 \sigma_{cyc}^2)]$.
This model gave a good fit with a $\chi^2/d.o.f. = 453/421$. 
The best fit parameters are shown in Table 1; the observed spectrum 
and the residuals (in units of $\sigma$) are shown in Figure 4 (upper and 
middle panel, respectively).
With respect to this model residuals are still visible in the soft energy
range, below $\sim 1$ keV, and in the PDS range, between 40 and 60~keV. 
Since the latter is the range of energy in which we expect 
to find harmonics of the cyclotron line we first added a cyclotron line
fixing the energy at twice the fundamental: this did not improve the fit.
Therefore we let the cyclotron line energy vary. 
In this way we obtain a reduction of the $\chi^2$ to
$442/418$. An F-test gives a probability of chance improvement of 1\%. 
The energy of the second absorption line is $51^{+4}_{-3}$~keV, its
depth is $0.9^{+0.1}_{-0.4}$, and its width is $3.3^{+5}_{-1}$~keV. 
The residuals with respect to this model, containing two cyclotron lines, 
are shown in Figure 4 (lower panel).
As it is visible in this figure, some residuals remain in the soft energy
range, below 1 keV.  The addition of a soft blackbody to the model reported 
in Table 1 improves the fit, giving $\chi^2/d.o.f. = 433/419$, corresponding 
to a probability of chance improvement of $\sim 8.4 \times 10^{-3}$. 
The blackbody has a temperature of $0.08 \pm 0.04$ keV and a luminosity of 
$\sim 1.6 \times 10^{37}$ ergs/s (although this last parameter is not well
constrained by the data,
and ranges between $1.3 \times 10^{36}$ ergs/s and $7.6 \times 10^{39}$ 
ergs/s).
Moreover, the spectral model for the soft excess is not univocally
determined: a bremsstrahlung component with a temperature of $0.10 \pm 0.07$
keV gives an equally good fit.

We also tried the so-called NPEX model to fit the continuum. This model
has been successfully used to fit the continuum of a number of HMXBs 
(Mihara 1995; Makishima et al. 1999) and consists of a negative plus a 
positive power law with a common exponential cutoff. It approximates the 
the spectrum produced by unsaturated thermal Comptonization in a
plasma of temperature T.  Using NPEX together with the absorption cyclotron 
feature, modelled by a pseudo--Lorentzian shape, and the iron emission line 
we obtain a negative photon index of $0.59 \pm 0.07$ and a temperature of 
$4.93 \pm 0.07$ keV, with the positive photon index fixed to 2.
In this case we obtained a $\chi^2/d.o.f. = 463/419$,
that is not better than the previous model (cf.\ Tab.~1). The residuals
in the soft band (below 1 keV) and in the hard range (around 50 keV) are 
still visible.

\subsection{Pulse Phase Resolved Spectra}

The hardness ratio shown in Figure~3 indicates the presence of significant
variations of the spectrum with the pulse phase. To investigate these
variations we have divided the pulse profile in 4 phase intervals: 0--0.5,
corresponding to the secondary peak; 0.5--0.65, corresponding to the rising
phase of the primary peak; 0.65--0.75, corresponding to the maximum of the
primary peak; 0.75--1, corresponding to the descent phase of the primary 
peak (see Fig.~2).  We produced energy spectra corresponding to each of 
these phase intervals, which were fitted by the same model used to fit the
pulse-phase averaged spectrum, i.e.\ a power law with high energy cutoff, the
iron emission line, and the absorption cyclotron line. The results are 
reported in Table~1.  

The energy of the cyclotron line varies
between 20.8~keV at the primary-peak maximum and 22.8~keV in the rising
phase of the primary peak, although these variations are not highly 
significant when compared to the associated error bars. More significant
variations are observed in the depth and width of the cyclotron line.
In particular the depth is larger when the flux is lower, while
the width of the line reaches its maximum and minimum values in the rising
and descent phase of the primary peak, respectively. 
Other significant variations with the pulse phase are in the shape of the
continuum. In particular the power law is flatter in correspondence
of the primary and secondary peaks, while it is steeper at the other phase
intervals. The energy of the cutoff seems also to be lower in correspondence
of the primary and secondary peaks, while no clear variation is present in 
the {\it e}-folding energy. Finally the iron emission line is compatible with
being unchanged along the pulse profile, although it seems to be weaker 
in the rising phase and at the maximum of the primary peak.
No statistically significant residuals are observed in the energy range 
between 40 and 50 keV, and therefore no second cyclotron harmonic is 
required to fit these spectra, although this could be due to the lower 
statistics of the phase resolved spectra with respect to the averaged 
spectrum.  Residuals below 1 keV are visible in the spectra at phases
0--0.5 and 0.75--1. In these cases the addition of a soft blackbody
gives a probability of chance improvement of the fit of 0.02\% and 0.08\%,
respectively. The temperatures of this component were $0.032 \pm 0.006$ 
and $0.10 \pm 0.04$ keV, respectively.

To better investigate the variation of the power-law photon index with the
pulse phase, we divided the pulse profile in 10 intervals and produced 
energy spectra from LECS and MECS data (for which the statistics is the 
highest) for each for these intervals. These spectra were fitted by a
power law and a gaussian emission line, with photoelectric absorption.  
The measured photon index is
reported versus the corresponding phase interval in Figure 5 (lower panel)
together with the corresponding flux in the 2--10~keV energy band (upper
panel).  As it can be seen from this figure, there is a clear anti-correlation
between the photon index and the flux, with the spectrum being flatter
in correspondence of higher flux levels.

\subsection{Spectral Evolution during the Eclipse}

The eclipse of the X-ray source is clearly visible at the beginning of the
BeppoSAX observation (see Fig.~1).
To study the spectral evolution during the eclipse, we considered the 
MECS spectra corresponding to the following three intervals. 
The first interval (the ingress) corresponds to the first 20 ks of the 
light curve, during which the soft (0.1--2.5 keV) emission was already 
at its minimum, while some hard emission is clearly visible in the MECS 
and PDS light curves. The second interval (the deep eclipse)
corresponds to the 50 ks in which the flux of the source was at its minimum
in all the energy bands. The third interval (the egress) corresponds to the
successive 20 ks in which the source gradually came back to its normal
flux level. The spectra were rebinned in order to have at least 30 counts
in each energy bin.

These spectra cannot be fitted by a simple power law with photoelectric 
absorption: the main features in the residuals are a soft excess below 
$\sim 3$ keV and a prominent excess between 6 and 7 keV.  
We therefore fitted these 
spectra with the same model used for the averaged spectrum 
out of eclipse (see Tab.~1, i.e.\ a power law, a Gaussian emission line 
and the Galactic photoelectric absorption) multiplied by a partial covering 
component. In this model, the Galactic hydrogen column 
$N_H$, the power-law photon index and the emission line parameters were fixed
to the values found for the averaged out-of-eclipse spectrum, while the
partial covering parameters, i.e.\ the equivalent hydrogen column $N_{Hpc}$
and the covering fraction $f_{pc}$, as well as the power-law normalization
were considered free parameters in the fit. This model could well fit the 
ingress and egress spectra, but gave a poor fit to the deep-eclipse spectrum,
for which we obtained a $\chi^2/d.o.f. = 235/58$. The main feature in the 
residuals was a soft excess below $\sim 3$~keV. We obtained a significant
improvement of the fit allowing the power-law photon index to vary.
The best fit value of this parameter was $\sim 2.6$, and the corresponding
$\chi^2/d.o.f.$ was $74/57$. The results of these fits are reported in 
Table~2 and the unfolded spectra are shown in Figure~6.

We also tried to fit these spectra with a more complex model for the 
continuum, i.e.\ the model that has been suggested to explain the emission 
during the eclipse in Cen~X--3 (e.g.\ Nagase et al. 1992; 
Ebisawa et al. 1996).  
This model consists of three components: the direct component 
from the neutron star, the electron-scattered component from the 
circumstellar matter and the dust-scattered component from the interstellar
dust. The first two have the same spectral shape, that we modelled with 
power laws with photon indices fixed to 1, but different normalizations and
absorption columns. The third one is softer by approximately $E^{-2}$ and
was modelled by a power law with photon index fixed to 3.
With this model we can obtain a good description of the deep-eclipse spectrum
if we also add a Gaussian emission line at 6.66 keV (with $\sigma \sim 0.17$
and equivalent width $\sim 4$ keV), giving a $\chi^2/d.o.f. = 58/53$.
For the other two spectra (eclipse ingress and egress) we do not obtain a 
stable fit with this model and the spectral parameters are not constrained. 
This is due to the fact that one of these components is indeed not required
by the data and could be eliminated without affecting the results.
Therefore we prefer the previous spectral deconvolution using the partial
covering, which has less free parameters and gives stable fits to all the
spectra. This model can well describe both the evolution of the soft part
of the spectra and the prominent feature at 6--7 keV, without the addition 
of extra emission components.

\section{Discussion and Conclusions}

We performed temporal and broad band (0.1-100~keV) spectral analysis on the
out-of-eclipse and eclipse data of 4U~1538--52 observed by BeppoSAX NFIs. 
We obtained a new measurement of the spin period 
$P_{\rm spin} = 528.24 \pm 0.01$ s, which 
indicates that the neutron star is still spinning-up. The pulse
profile is double peaked in the MECS energy 
range (1.8-10.5~keV), with a main pulse and a secondary one. The secondary
peak is much fainter in the high energy range (15-100~keV) and  
absent in the soft energy range (0.1-1.8~keV). 

The broad band energy spectrum is well fitted by a power-law with high energy
cutoff continuum (the typical continuum of the X-ray pulsators), with low 
energy absorption by cold matter, an emission line due to fluorescence from 
iron in low ionization stages, and an absorption feature around 20~keV, 
interpreted as due to cyclotron scattering (in agreement with {\it e.g.} 
Clark {\it et al.} 1990). 
The cyclotron line parameters are similar to those  
measured by Ginga (Clark {\it et al.} 1990; Makishima et al. 1999), with 
small differences due to the different models
(Gaussian vs. pseudo--Lorentzian) used to fit the cyclotron line.   
A turnover of the spectrum above the cyclotron line energy was previously
interpreted as due to the presence of the second cyclotron harmonic or 
a high energy thermal cutoff (Clark {\it et al.} 1990). 
Makishima et al. (1999),
using the NPEX model, interpreted this turnover in the spectrum as due
to a thermal cutoff, although they fitted the spectrum in a relatively 
narrow energy range (2--60 keV) and obtained a $\chi^2/d.o.f. 
\simeq 46/29$.  
The broad band capabilities of the BeppoSAX NFIs (0.1--100 keV) have 
unambiguously demonstrated that the claimed second harmonic present at the 
edge of the energy range of the {\it Ginga} LAC actually is an exponential
cutoff, that we also interpret as thermal with associated temperature given 
by the e-folding energy of $\sim 10$~keV. If this interpretation is
correct we expect a Gaussian thermal broadening of the cyclotron 
absorption feature according to the formula: 
$\sigma_{\rm cyc} / E_{\rm cyc} = (k T / m_{\rm e} c^2)^{1/2}$ 
(Rybicki \& Lightman 1979), where $m_{\rm e} c^2$ is the rest energy
of the electron. For $E_{\rm cyc} = 21$~keV, and $k T = 10$~keV, we found
$\sigma = 2.9$~keV, in agreement with the measured broadening
of $\sigma = 3.4$~keV.  

The thermal nature of the broadening of the cyclotron absorption
feature seems to be the more natural interpretation for the width of
these lines, although different interpretations have also been proposed.  
It seems possible that variations of the field strength along the
accretion column could originate the variations of the broadening
observed in the pulse phase resolved spectroscopy performed on some
sources. Indeed, Bulik {\it et al.} (1995) claim that multiple field
values are needed to explain the pulse phase resolved spectra of
{\it Ginga} data of 4U~1538--52. Although this is still a viable
possibility, we want to stress that, thanks to the broad band
capabilities of the BeppoSAX satellite, in an increasing number of
X-ray pulsators the shape of the high energy continuum (underlying the
absorption cyclotron lines) has been determined quite accurately.  In
particular the presence of a power law with an exponential cutoff
seems a common feature among these sources. If a multiplicative
gaussian absorption feature is used to fit the cyclotron lines a good
fit has been obtained in several cases. 
We note that a multiplicative Gaussian shape (that is also used in this 
paper) can better describe the effects of thermal broadening of the
cyclotron feature (as discussed in Burderi et al. 2000).
As it can be seen in Dal Fiume et al. (1999), there is a general agreement
between the measured width of cyclotron lines in several sources and the 
predicted value of the width in the hypothesis of thermal broadening.

Another absorption feature might be present in the spectrum of
4U~1538--52 around 50~keV (at 99\% confidence level). 
The presence of variations of the pulse profile at energies below
and above 50 keV could provide a confirmation of the presence of a
cyclotron feature at this energy. Unfortunately, although the primary
peak is still observed in the folded light curve at energies above 50 keV, 
the reduced statistics prevent us to see significant variations of the 
pulse profile at high energy.  
If confirmed, this absorption feature might be 
interpreted as the second harmonic of the 20~keV
cyclotron line. However, the energy of this second feature
($51^{+4}_{-3}$~keV) does not seem compatible with being double than
that of the fundamental ($21.1 \pm 0.2$~keV).  This discrepancy could
be explained by considering that the optical depths of the fundamental
and second harmonic are different, implying that the two lines could
form at different heights, where the dipolar magnetic field of the
neutron star has different intensities.  In particular the
fundamental, with a larger cross section, should form in the upper
atmosphere where the magnetic field is weaker. We calculate that a
difference in the heights of $\sim 0.07$ neutron star radii is
sufficient to explain the difference in the cyclotron energies.
Another possibility is that we see two different lines coming from the
two magnetic poles. A displacement of the magnetic dipole momentum by
about 0.15 neutron star radii from the center (as already detected in
other X-ray pulsators, {\it e.g.} Cen X-3, Burderi {\it et. al.},
2000) is sufficient to explain the difference of energy of the two
lines produced at the two magnetic poles. In this case we expect a
phase dependence of the strength of these two lines.  In principle
phase resolved spectral analysis could address this question although
the low statistics at the energies of the second line prevents any
conclusion with the present data set.  Further observations are needed
to confirm the presence of the second harmonic and to address this
interesting question.

We also performed a pulse-phase resolved spectral analysis on these
data, dividing the pulse profile into four intervals, corresponding to
the secondary peak and the rising, maximum and descent phases of the 
primary peak.  The cyclotron line energy seems to vary with the pulse
phase, reaching its maximum and minimum values in the rising phase and 
at the maximum of the primary peak, respectively, although the large 
error bars prevent any firm conclusion based on these data.
The line depth is higher at the secondary-peak phase, while the width 
reaches its maximum and minimum values at the rising and descent phases 
of the primary peak, respectively.  These variations are in general 
agreement with what is observed, with higher statistical significance, 
in the {\it Ginga} data of 4U~1538--52 (Clark et al. 1990; Bulik et al. 1992).
Significant variations with the pulse phase are also observed in the 
continuum model. In particular the power law is flatter at the phases 
of the two peaks, and the corresponding values of the cutoff energy
tend to be lower than in the other phase intervals.  This behavior
of the photon index is better seen in Figure 5, where we show the results
of the fit of ten phase intervals in the LECS and MECS ranges. These
results are similar with previous results from EXOSAT (Robba et al. 1992)
and Ginga (Clark et al. 1990), and hence they probably reflect true
variations of the Compton $y$ parameter and the temperature of the
observed region in the accretion column with the pulse phase.

A soft blackbody component might also be present in the pulse-phase 
averaged spectrum as well as in the spectra corresponding to phase
intervals 0--0.5 and 0.75--1. The temperature of this component in the
pulse-phase averaged spectrum is $\sim 0.08$ keV, corresponding to
an unabsorbed luminosity of $\sim 1.6 \times 10^{37}$ ergs/s. The temperature 
seems to vary between $\sim 0.03$ keV at phases 0--0.5 and $\sim 0.1$ keV
at phases 0.75--1.  The radius of the spherical emission region of this
component is between 400 and $2 \times 10^4$ km, well compatible with
the magnetospheric radius of this source. In fact, considering a magnetic
field of $1.8 (1+z)\times 10^{12}$ Gauss (where $z$ is the gravitational
redshift) and a bolometric luminosity of $\sim 2.2 \times 10^{37}$ ergs/s,
we find 
$R_{\rm M} \sim 5 \times 10^3 \phi (1+z)^{4/7}$ km, where $\phi$ is
a correction factor which, in the case of disk accretion, is between 0.3 and
0.5 (see e.g.\ Burderi et al. 1998; Ghosh \& Lamb 1991).
Therefore this soft component can be emitted by the inner accretion
disk at the magnetospheric radius or by matter at the magnetosphere
(as in the case of Cen X--3, Burderi et al. 2000).

Finally,
we analysed eclipse spectra during the ingress, the deep eclipse and the
egress. The best fit to these spectra was obtained adopting the same model
used for the out-of-eclipse spectrum multiplied by a partial covering.
In this model the spectral evolution during
the eclipse is explained by a progressive covering of the primary spectrum
(or the part of it that is scattered into the line of sight), with both
$N_{Hpc}$ and $f_{pc}$ increasing towards the deep eclipse.  
This model can fit both a soft excess and a prominent feature between
6 and 7 keV without the addition of extra emission components. In the deep
eclipse we also find a softening of the power-law component. This could be
caused by the increased contribution of the dust-scattered component, which
has a softer spectrum (by $\sim E^{-2}$) relative to the incident spectrum
(e.g.\ Day \& Tennant 1991).

\acknowledgments
This work was supported by the Italian Space Agency (ASI) and 
by the Ministero della Ricerca Scientifica e Tecnologica (MURST).


\clearpage

\begin{table}[h]
\small
\begin{center}
\caption{Results of the fit of the pulse phase selected and averaged spectra
in the energy range 0.1-100~keV. 
Uncertainties are at the 90\% confidence level for a single parameter. 
The power-law normalization, $N$, is in unit of ph~keV$^{-1}$ cm$^{-2}$ 
s$^{-1}$ at 1~keV. The Gaussian emission line intensity, $I_{\rm Fe}$, is 
in units of ph cm$^{-2}$ s$^{-1}$.
The flux, in units of $10^{-10}$ ergs~cm$^{-2}$~s$^{-1}$, is calculated 
in the 0.1--100 keV energy range.}
\label{table1}
\vspace{0.5cm}
\begin{tabular}{l|c|c|c|c|c} \hline \hline
Phase  & $0-0.5$         & $0.5-0.65$    & $0.65-0.75$ & $0.75-1$ & Averaged 
\\ \hline
$N_H$ ($\times 10^{22}$~cm$^{-2}$) & $1.70 \pm 0.07$ & $1.8 \pm 0.1$ & 
$1.4 \pm 0.1$ & $1.58 \pm 0.08$ & $1.63 \pm 0.04$ \\
       &                 &		 & 		 &	& \\
P.I.   & $1.06 \pm 0.02$ & $1.39 \pm 0.04$ & $0.97 \pm 0.03$ & 
$1.12 \pm 0.02$ & $1.12 \pm 0.01$ \\
N $(\times 10^{-2})$  & $2.6 \pm 0.1$ & $6.6 \pm 0.4$ &
$4.3 \pm 0.2$ & $4.8 \pm 0.2$ & $3.8 \pm 0.8$ \\  
$E_{\rm cut}$ (keV)  & $16 \pm 1$ & $25 \pm 3$ & $14 \pm 1$ & $17 \pm 1$ &
$16.4\pm 0.7$ \\    
$E_{\rm fold}$ (keV) & $7.8 \pm 0.8$ & $10.3 \pm 2$ & $11.9 \pm 0.9$ & 
$9.3 \pm 0.7$ & $10.0\pm 0.5$  \\ 
       &                 &  		 &   		 &  	&  \\
$E_{\rm cyc}$ (keV) & $21.1 \pm 0.3$ & $22.8 \pm 1$ & $20.8 \pm 0.6$ & 
$21.5 \pm 0.5$ & $21.1 \pm 0.2$  \\
Depth               & $0.69  \pm 0.06$ & $0.56 \pm 0.02$ & $0.36 \pm 0.07$ & 
$0.37 \pm 0.07$ & $0.49  \pm 0.04$ \\
$\sigma_{\rm cyc}$ (keV) & $3.7 \pm 0.4$ & $5.6 \pm 1.4$ & $3.6 \pm 0.8$ & 
$2.5_{-0.8}^{+0.6}$ & $3.4 \pm 0.3$ \\
      &              &                 & 	     &		    & \\
$E_{\rm Fe}$ (keV)  & $6.44 \pm 0.06$ & 6.4 (frozen) & 6.4 (frozen) & 
$6.38 \pm 0.09$ & $6.37 \pm 0.05$ \\ 
$\sigma_{\rm Fe}$ (keV)  & 0.0 (frozen) & 0.0 (frozen) & 0.0 (frozen) & 
0.0 (frozen) & 0.0 (frozen) \\
$I_{\rm Fe}$ $(\times 10^{-4})$ & $3 \pm 1$ & $< 3$ & 
$< 2.3$ & $2.8_{-0.8}^{+1}$ & $2.5 \pm 0.5$ \\
EW$_{\rm Fe}$ (eV)  & 85 & $< 43$ & $< 52$ &  57 & 53   \\
                    &    &    &    &      &      \\
Flux                & 6.46 & 9.97 & 16.4 & 12.5 & 9.60  \\
$\chi^2$/d.o.f.     & 483/495 & 391/394 & 374/371 & 504/467 & 453/421  \\
\hline
\end{tabular}
\end{center}

\small
\begin{center}
\caption{Results of the fit of the spectra during the eclipse
in the energy range 1.8-10~keV. Uncertainties are 
at the 90\% confidence level for a single parameter. 
The units of the spectral parameters are as defined in Table~1.}
\label{table1}
\vspace{0.5cm}
\begin{tabular}{l|c|c|c} \hline \hline
                & Ingress & Eclipse & Egress \\ \hline
$N_H$	& 1.63 (frozen) & 1.63 (frozen) & 1.63 (frozen) \\
P.I.	& 1.12 (frozen) & $2.68^{+0.2}_{-0.1}$ & 1.12 (frozen) \\
Norm    & $(2.56 \pm 0.08) \times 10^{-2}$ & $0.3^{+0.2}_{-0.1}$ &
	  $(3.5 \pm 0.1) \times 10^{-2}$ \\
$N_{Hpc}$ & $17 \pm 1$  & $140^{+30}_{-20}$ & $8 \pm 1$ \\
$f_{pc}$  & $0.916 \pm 0.009$ & $0.982^{+0.007}_{-0.010}$ & 
	    $0.80 \pm 0.03$ \\
$\chi^2$/d.o.f.     & 150/154 & 74/57 & 174/167   \\
\hline
\end{tabular}
\end{center}
\end{table} 

\newpage

\begin{figure}
\plotone{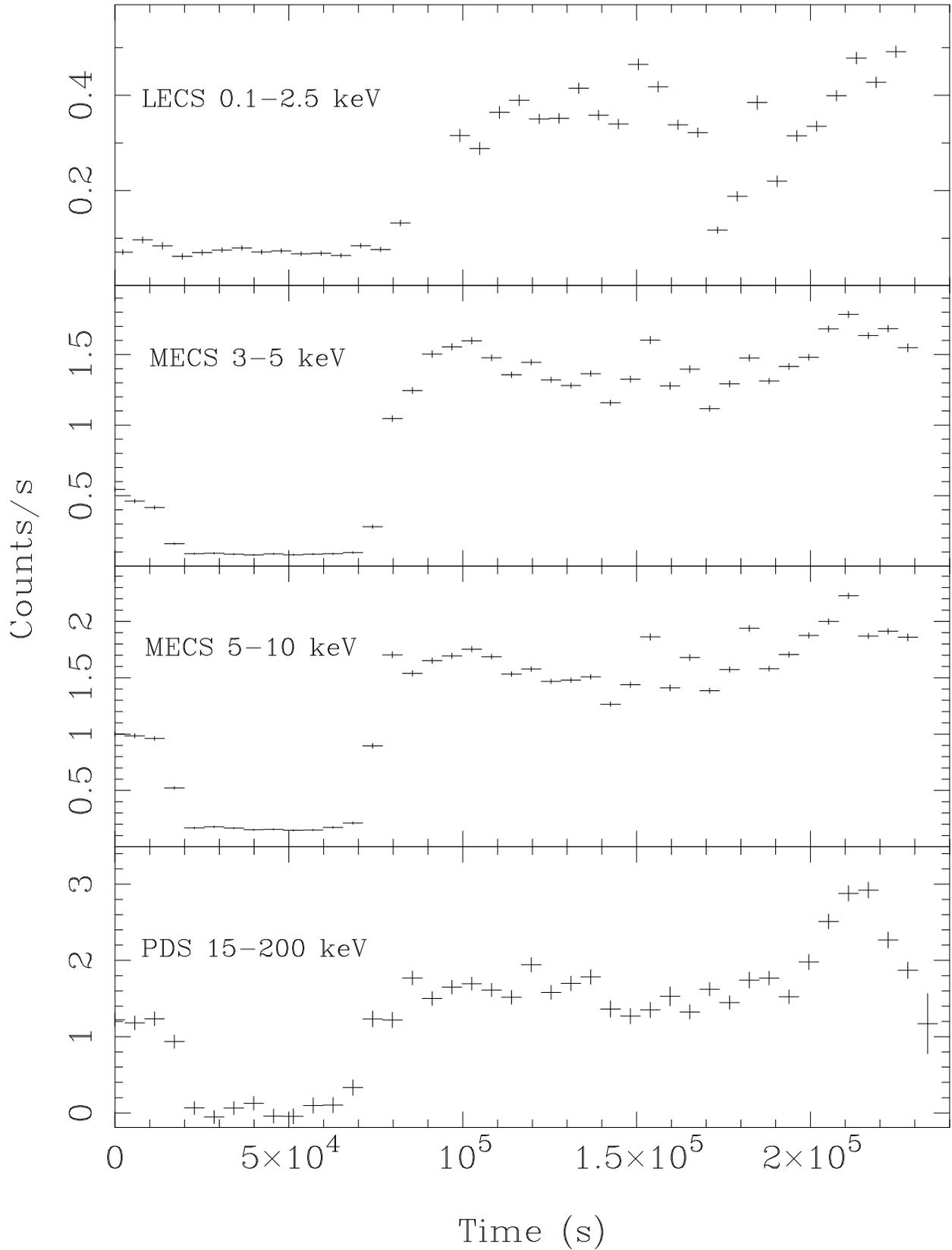}
\caption{\label{fig:fig1} 4U~1538--52 light curve, binned at 5700 s, in four 
different energy bands: 
0.1--2.5~keV (LECS data), 3--5~keV (MECS data), 5--10~keV (MECS data),
15--200~keV (PDS data).}
\end{figure}

\begin{figure}
\plotone{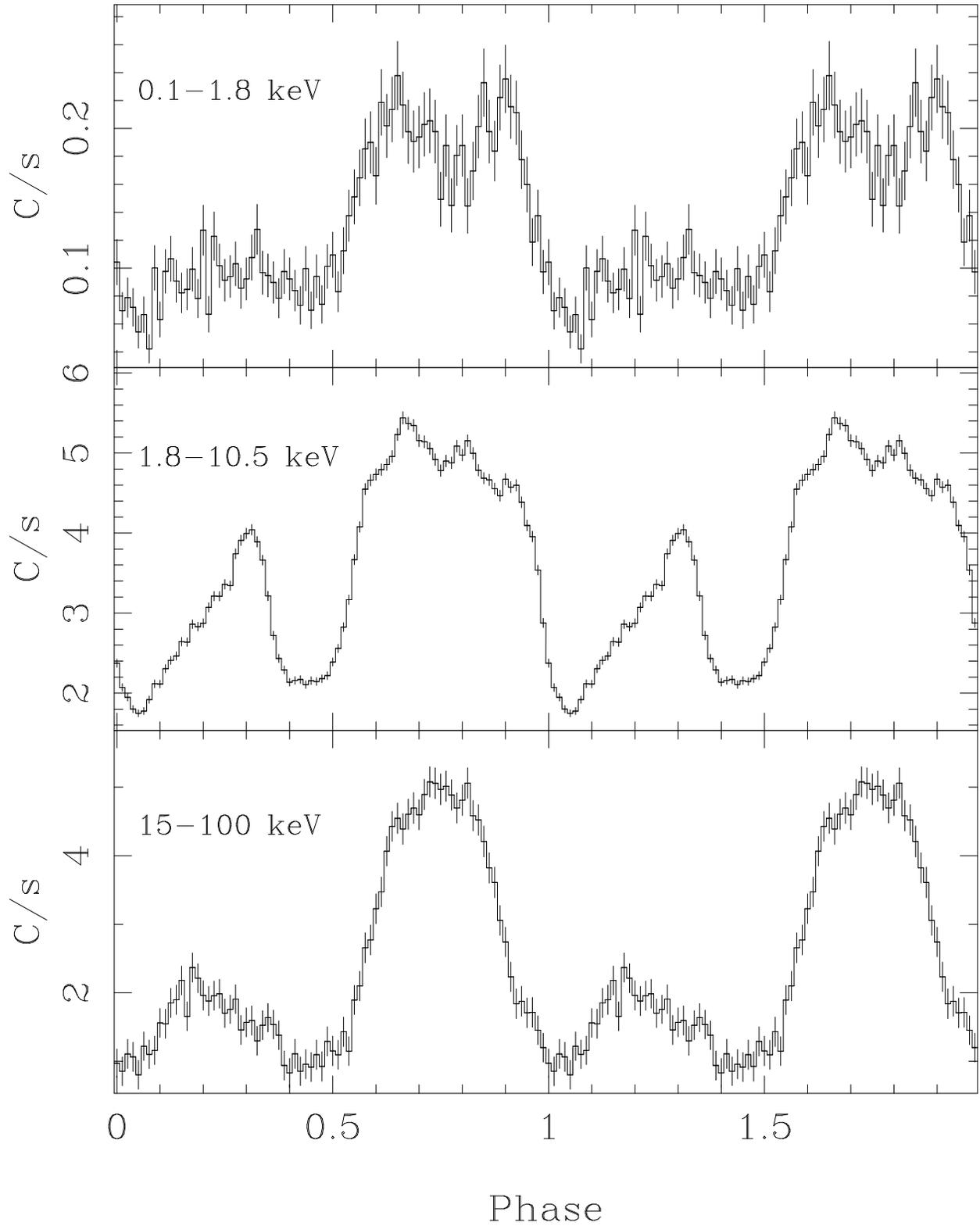}
\caption{\label{fig:fig2} Pulse profiles during the post-egress phase in different 
energy bands: 0.1-1.8~keV (LECS, upper panel), 1.8-10.5~keV (MECS, 
middle panel), 15-100~keV (PDS, lower panel).}
\end{figure}

\begin{figure}
\plotone{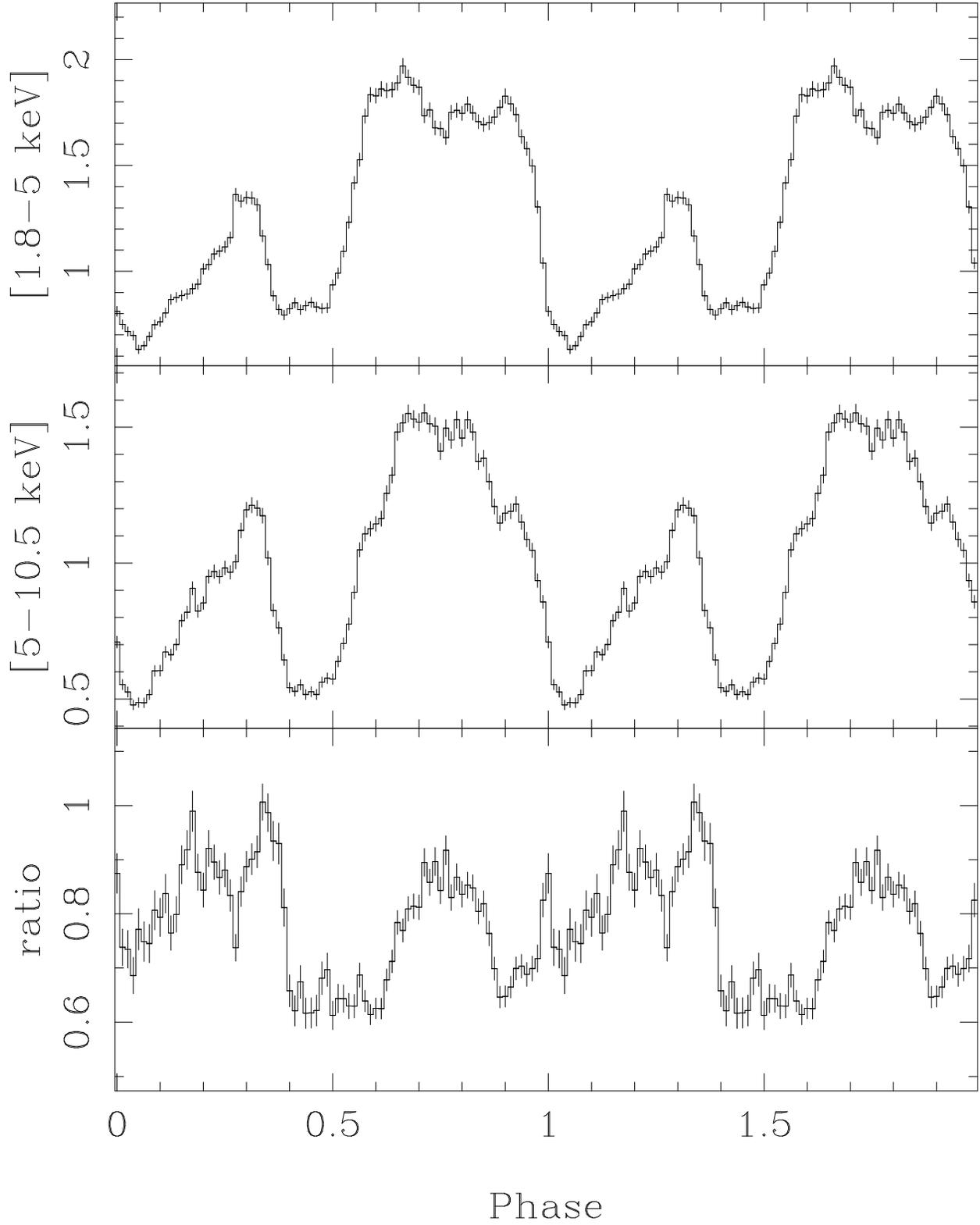}
\caption{\label{fig:fig3} MECS folded light curves in the energy ranges
1.8-5~keV (upper panel) and 5-10.5~keV (middle panel), and their hardness 
ratio (lower panel). }
\end{figure}

\begin{figure}
\plotone{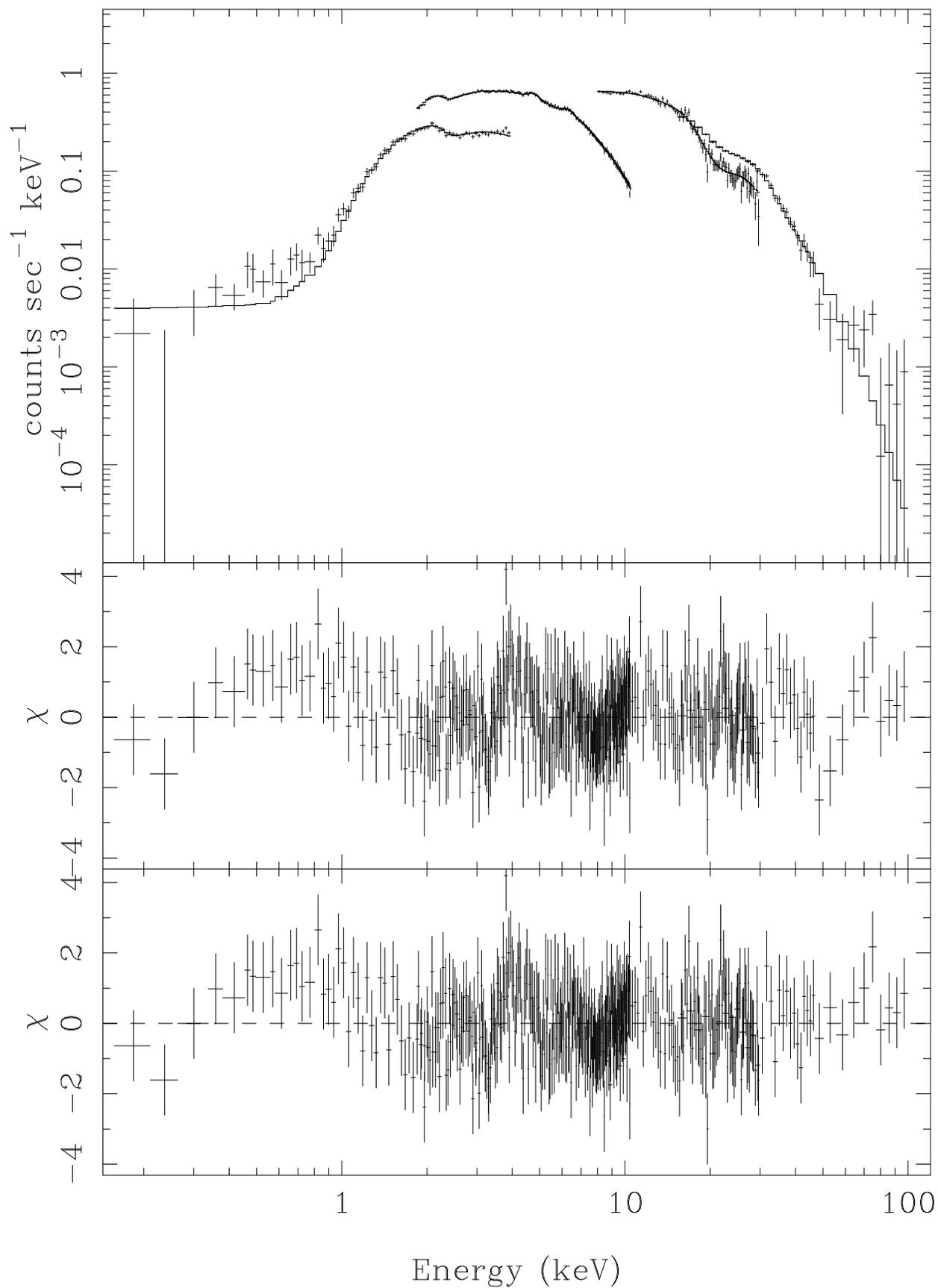}
\caption{\label{fig:fig4} Broad band spectrum of 4U~1538--52 (upper panel),
residuals, in units of $\sigma$, with respect to the shown in Table 1
(middle panel), and residuals, in units of $\sigma$, with respect to 
the model with two cyclotron lines (lower panel).}
\end{figure}

\begin{figure}
\plotone{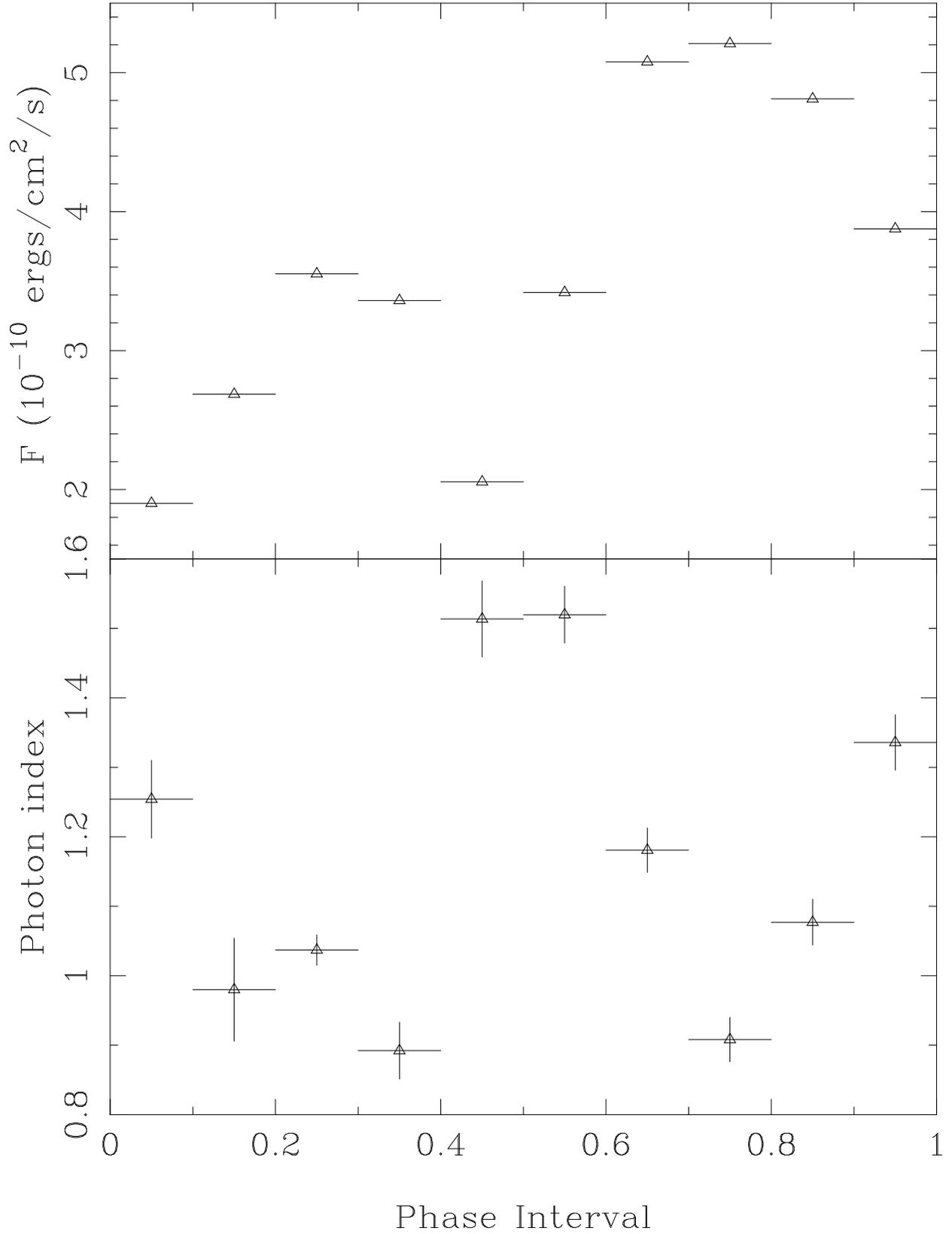}
\caption{\label{fig:fig5} Measured photon index (lower panel)
and the corresponding flux in the 2--10~keV energy band (upper
panel) reported versus the corresponding phase interval.}
\end{figure}

\begin{figure}
\plotone{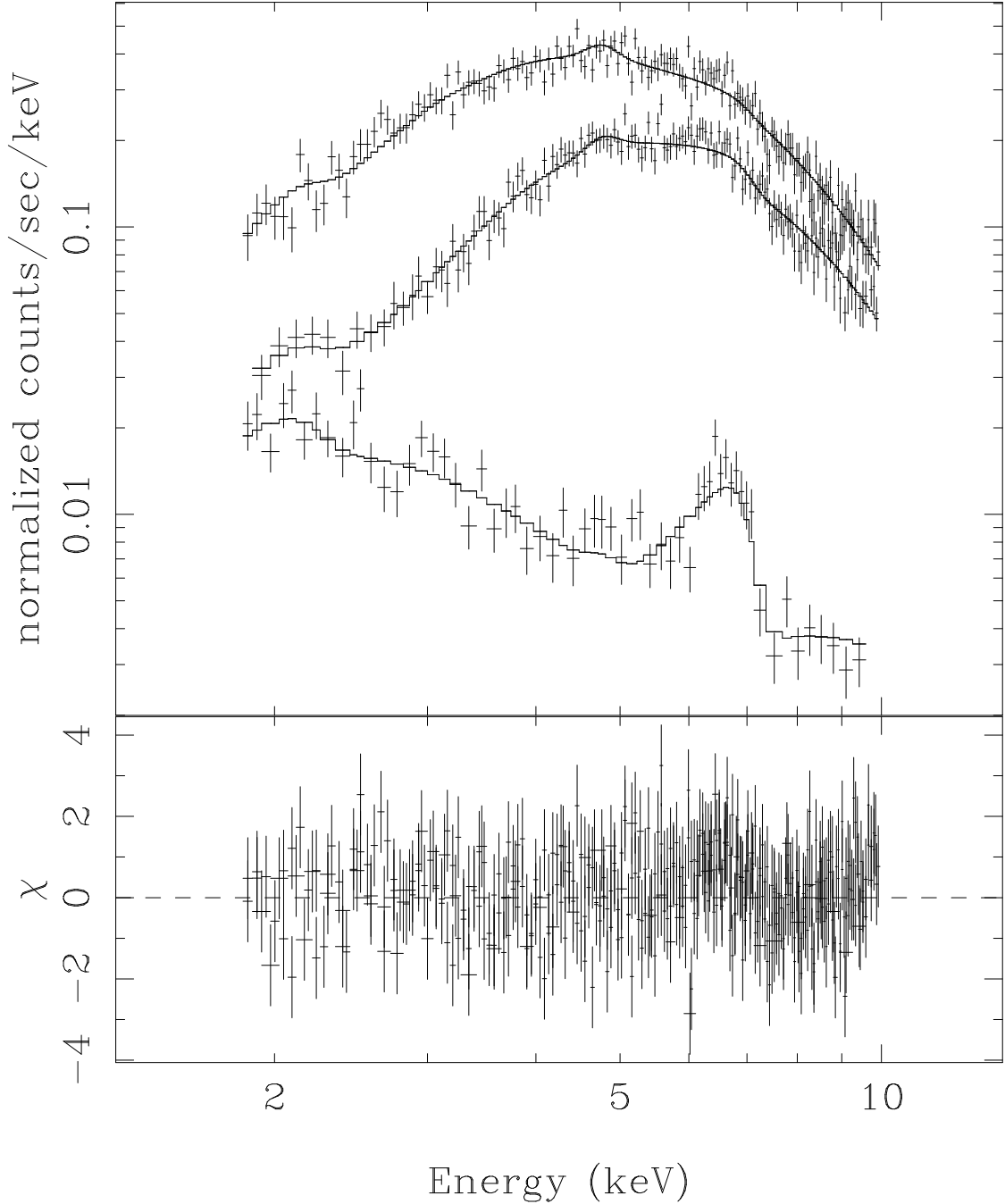}
\caption{\label{fig:fig6} Top panel: spectra of 4U~1538--52 during the ingress 
(middle), the deep eclipse (bottom) and the egress (top), together with the 
spectral model used to fit these spectra (see Tab.~2). Bottom panel: 
corresponding residuals in units of $\sigma$.}
\end{figure}

\end{document}